\documentclass{article}

\usepackage{arxiv}

\usepackage[T1]{fontenc}
\usepackage{cite}
\usepackage{amsmath,amssymb,amsfonts}
\usepackage{algorithmic}
\usepackage{graphicx}
\usepackage{textcomp}
\usepackage{xcolor}
\usepackage{hyperref}
\usepackage{siunitx}
\usepackage[capitalise]{cleveref}

\def\BibTeX{{\rm B\kern-.05em{\sc i\kern-.025em b}\kern-.08em
    T\kern-.1667em\lower.7ex\hbox{E}\kern-.125emX}}
    
\begin{document}
\title{PyPSA meets Africa: Developing an open source electricity network model of the African continent}

\author{Desen Kirli \\
\textit{School of Engineering} \\
\textit{University of Edinburgh}\\
Edinburgh, Scotland, United Kingdom \\
\href{mailto:desen.kirli@ed.ac.uk}{desen.kirli@ed.ac.uk} 
\href{https://orcid.org/0000-0001-7596-0944}{\includegraphics[scale=0.06]{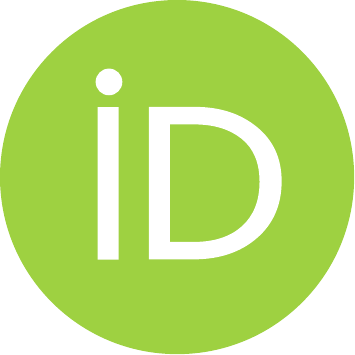}} \\
\And
Johannes Hampp\\
\textit{Center for International Development} \\
\textit{and Environmental Research} \\
\textit{Justus-Liebig-University Giessen}\\
Giessen, Germany \\
\href{mailto:johannes.hampp@zeu.uni-giessen.de}{johannes.hampp@zeu.uni-giessen.de}
\href{https://orcid.org/0000-0002-1776-116X}{\includegraphics[scale=0.06]{orcid.pdf}}
\And
Koen van Greevenbroek\\
\textit{Department of Computer Science} \\
\textit{UiT The Arctic University of Norway}\\
Tromsø, Norway \\
\href{mailto:koen.v.greevenbroek@uit.no}{koen.v.greevenbroek@uit.no}
\href{https://orcid.org/0000-0002-6105-2846}{\includegraphics[scale=0.06]{orcid.pdf}}
\And
Rebecca Grant\\
\textit{School of Geosciences} \\
\textit{University of Edinburgh}\\
Edinburgh, Scotland, United Kingdom \\
\href{mailto:r.grant-20@sms.ed.ac.uk}{r.grant-20@sms.ed.ac.uk} \\
\And
Matin Mahmood\\
\textit{School of Informatics} \\
\textit{University of Edinburgh}\\
Edinburgh, Scotland, United Kingdom \\
\href{mailto:M.Mahmood-3@sms.ed.ac.uk}{m.mahmood-3@sms.ed.ac.uk} \\
\And
Maximilian Parzen\\
\textit{School of Engineering} \\
\textit{University of Edinburgh}\\
Edinburgh, Scotland, United Kingdom \\
\href{mailto:max.parzen@ed.ac.uk}{max.parzen@ed.ac.uk}
\href{https://orcid.org/0000-0002-4390-0063}{\includegraphics[scale=0.06]{orcid.pdf}}
\And
Aristides Kiprakis\\
\textit{School of Engineering} \\
\textit{University of Edinburgh}\\
Edinburgh, Scotland, United Kingdom \\
\href{mailto:kiprakis@ed.ac.uk}{kiprakis@ed.ac.uk}
\href{https://orcid.org/0000-0001-7596-0944}{\includegraphics[scale=0.06]{orcid.pdf}}
}

\maketitle

\begin{abstract} 
Electricity network modelling and grid simulations form a key enabling element for the integration of newer and cleaner technologies such as renewable energy generation and electric vehicles into the existing grid and energy system infrastructure. This paper reviews the models of the African electricity systems and highlights the gaps in the open model landscape. Using \emph{PyPSA} (an open Power System Analysis package), the paper outlines the pathway to a fully open model and data to increase the transparency in the African electricity system planning.
Optimisation and modelling can reveal viable pathways to a sustainable energy system, aiding strategic planning for upgrades and policy-making for accelerated integration of renewable energy generation and smart grid technologies such as battery storage in Africa.
\end{abstract}


\section{Introduction}
Open models and open-modelling initiatives bridge the gap between policy-making and long-term planning performed by researchers. For developing countries, including but not limited to the ones located in the African continent, it is essential to simulate long-term energy generation and consumption scenarios, considering economic viability, risk and rate of return as these are key for investments in rural electrification and integration of renewable energy sources (RES).

Existing open models and tools targeting Africa focus largely on regional electrification pathways and RES integration, but there is also recent development in continent-wide models. We review relevant open modelling efforts, and find a need for a high resolution long-term investment optimisation model for Africa.

The main contributions of this paper are outlined below:
\begin{itemize}
    \item A review of the existing open energy systems models of Africa in both academic and non-academic context.
    \item A roadmap for a new open python-based investment and dispatch optimisation model for the African energy system, with a high spatial and temporal resolution and accurate modelling of the transmission grid.
\end{itemize}

The proposed model will enable long-term analysis of factors such as RES integration, transmission expansion, electric vehicles and smart grid technologies. 

Since May 2021, the model development started with more than 27 members associated with European and African leading institutions in research and academia. A first prototype is aimed to be made available by the end of 2021. 
 
More information and the current developments can be accessed on the project website \footnote{\url{https://pypsa-meets-africa.github.io/}} and documentation \footnote{\url{https://pypsa-meets-africa.readthedocs.io/en/latest/index.html}}.

\section{Motivation \& Review of the Existing Work and Initiatives}
In this section, the motivation for an open high-resolution energy model for Africa is outlined.

Following this, we summarise the most relevant non-academic projects that promote open energy modelling in the African continent. In the second subsection, we review the academic literature that features existing open models and simulation tools used for either the whole continent or smaller regions in Africa. Lastly, we conclude this section by highlighting the gap that our proposed model addresses.

\subsection{Motivation and current developments for energy planning in Africa}
The main motivation behind most African energy models is to investigate and enable an increasing rate of electrification linked to the rapid urbanisation, economic development and population growth in the African continent \cite{energyprojections}.

In order to meet this predicted increase in electricity demand, respect limitations such as the greenhouse gas (GHG) emissions and help implement the UN's Sustainable Development Goals (SDGs), the Paris Agreement and Africa´s Agenda 2063, different assets (e.g. hydropower plants, wind turbines, batteries, etc.) and the African electricity transmission grid should be modelled. Additionally, other planned developments such as new transmission lines should be taken into account in long-term energy modelling.

\begin{figure}[h]
\centerline{\includegraphics[trim =  0cm 1.5cm 0cm 0cm, clip,width = 0.5\textwidth]{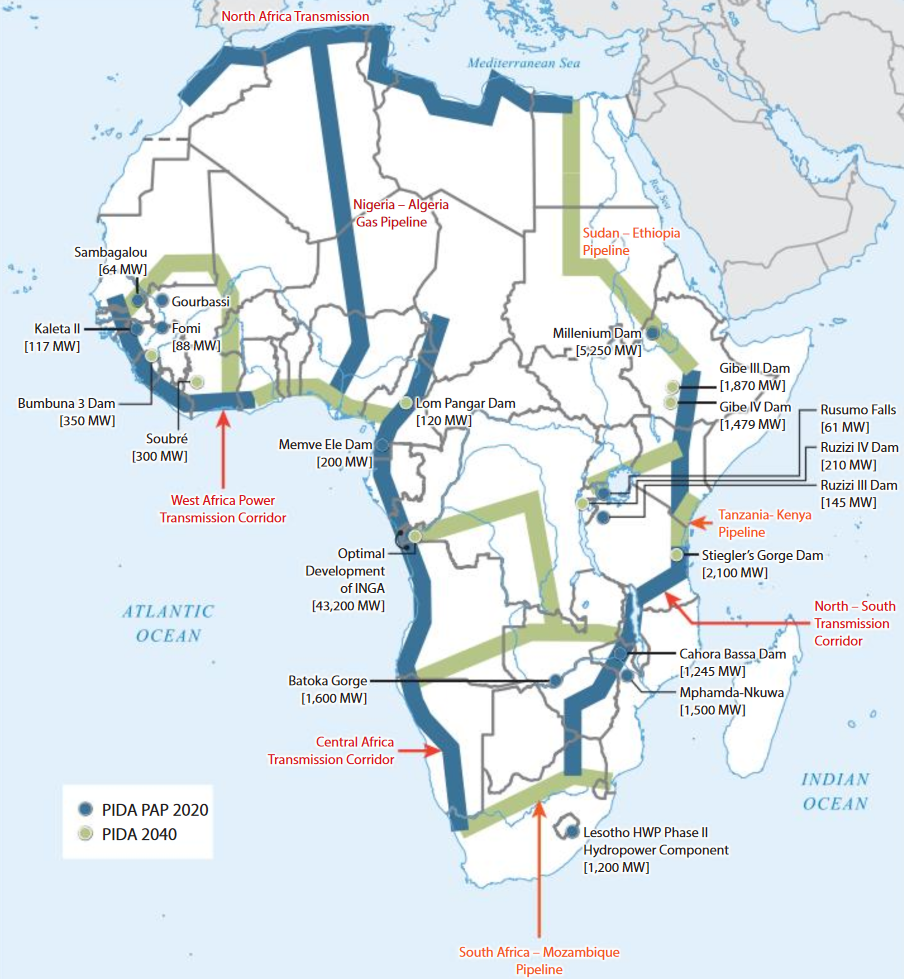}}
\caption{Interconnector expansion plans of the World Bank Programme for Infrastructure Development in Africa (PIDA) \cite{pida2015}. Blue and green lines refer to the 2020 Priority Action Plan and the 2040 plan respectively.}
\label{interconnector}
\end{figure}

\Cref{interconnector} shows planned interconnection projects between various countries in Africa. This is an example of many initiatives that attempt to increase the capacity and resilience of the African grid in order to prepare for the predicted increase in electricity demand. It is also a testament to the need for continent-level modelling.

In order to simulate the techno-economic benefits and costs of such planned developments and also test different RES policies, organisations such as the International Renewable Energy Agency (IRENA),
European Union (EU), International Energy Agency (IEA), United Nations Economic Commission for Africa (UNECA), African Climate Policy Centre (ACPC) and World Bank Group (WBG) provide funding, modelling expertise and also promote the use of open modelling tools \cite{IRENA, IEA, EMPA}.

To increase the contribution from RES, IRENA \cite{IRENA} established the Africa Clean Energy Corridor (ACEC) initiative. Using \emph{MESSAGE}, IRENA constructed several System Planning Test (SPLAT) models for 47 African countries. However, the SPLAT models focus on the analysis of zones that have a high solar and/or wind energy generation potential and have a resolution of 5 to 10-year time steps.
This tool is free for academic use only and the source code is not open, preventing external validation contributions. 

Another modelling initiative is the Energy Modelling Platform Africa (EMP-A) \cite{EMPA} which has many key contributors such as UNECA, ACPC, WBG, UK Department for International Development (DFID), United Nations Development Programme (UNDP) and more. They both promote and provide training for the use of two open packages, namely \emph{OSeMOSYS} and \emph{OnSSET}, for energy investment planning and geospatial electrification mapping applications, respectively.

For city-level modelling, the SAMSET (i.e. Supporting African Municipalities in Sustainable Energy Transitions) project uses \emph{LEAP} (i.e. Low Emissions Analysis Platform) which focuses on the sustainable energy transition of six African cities in Ghana, Uganda and South Africa \cite{SAMSET}.

Lastly,  the IEA's Energy Sub-Saharan Africa project, funded by the EU until 2023, aids Sub-Saharan African countries with their energy data management and long-term energy planning~\cite{IEA}. Although this is a great initiative as it pledges for open-access country-level energy data which can be used by our proposed model, there is no physics-informed power flow analysis involved.

To summarise, the main motivation behind continent-level open modelling is to perform long-term analysis of planned projects such as the interconnection example in \cref{interconnector} and simulate the inter-country power flows. Fuelled by the increase in economic development, urbanisation and population, the augmented electrical demand should be modelled and met from RES in order to comply with the UN's SDGs and ambitions to decrease emissions in the energy sector.

\subsection{Existing Research Initiatives}

Energy systems models in Africa have the same purpose as elsewhere: in a nutshell, exploring development and expansion pathways of energy systems.
One issue of particular importance to many regions in Africa is that of (rural) electrification, either by new grid connections or stand-alone options.
Some models are technologically detailed, being based on open frameworks such as \emph{OSeMOSYS}~\cite{howells-rogner-ea-2011}, \emph{Calliope}~\cite{pfenninger-pickering-2018} and \emph{PyPSA}~\cite{brown-horsch-ea-2018}.
Other models are based on extensive high-resolution data (demand, renewable generation potential, grid connections, etc.) and typically focus more exclusively on the question of electrification, e.g.\ models based on OnSSET~\cite{mentis-howells-ea-2017}.

These two characteristics do not split existing work neatly into two categories.
However, after the following brief literary review, we conclude that no model has been developed as of yet which is both technologically detailed \emph{and} models the entire African energy system at a high spatial resolution.
This is the gap we aim to fill, and which will lead to a better understanding of continent-wide energy development in Africa.

Mathematical modelling to map potential pathways for rural electrification is widespread, especially within the Global South~\cite{bertheau-cader-ea-2016,cader-blechinger-ea-2016,alfaro-miller-2021,rocco-fumagalli-ea-2021,anwar-deshmukh-ea-2020}.
Geospatial electrification analysis combines GIS analysis with least-cost technology planning approaches to map least-cost electrification options where electricity access is low~\cite{mentis-howells-ea-2017,trotter-cooper-ea-2019}.
This has been carried out at a country level in Sub-Saharan Africa including for Nigeria~\cite{mentis-siyal-ea-2017}, Ethiopia~\cite{mentis-andersson-ea-2016}, Kenya~\cite{mentis-howells-ea-2017}, Uganda~\cite{trotter-cooper-ea-2019}, in Malawi~\cite{korkovelos-khavari-ea-2019} and in Tanzania~\cite{menghwani-zerriffi-ea-2020}.
These case studies vary with respect to parameters used, though most include desired electricity target level and quality of access, population density, local grid connection status, existing electricity infrastructures, energy resource availability and cost of energy technologies in their assessments~\cite{menghwani-zerriffi-ea-2020}.
The \emph{OnSSET} model~\cite{mentis-howells-ea-2017,korkovelos-khavari-ea-2019} is especially popular as a spatial electrification analysis tool for Sub-Saharan Africa~\cite{rocco-fumagalli-ea-2021}.
Recently, it has been used to create high resolution electrification models for every African country.
The resulting pathways can be explored in the Global Electrification Platform\footnote{\url{https://electrifynow.energydata.info/}}.

At the same time, models have been developed for African regions which include transmission grid physics, optimal dispatch and storage operation, which \emph{OnSSET} does not.
These model the feasibility and operation of energy systems better, and allow us to study issues such as transmission bottlenecks and flexibility.
Good examples include a model of the South African energy system based on \emph{PyPSA}~\cite{horsch-calitz-2017} and a similar model (but focussed on concentrating solar power and nuclear power) based on Calliope~\cite{pfenninger-keirstead-2015}.
A significant development is \emph{TEMBA}, an \emph{OSeMOSYS}-based energy system model for the entire African continent~\cite{taliotis-shivakumar-ea-2016}.
It is used to study least-cost capacity expansion of generation and transmission on a continental scale.
Another recent addition is the \emph{Dispa-SET Africa} model~\cite{pavicevic-defelice-ea-2021}, a unit commitment and dispatch optimisation model.

\emph{TEMBA} is a large-scale capacity expansion model, but it is spatial resolution is limited to one node per country.
\emph{Dispa-SET Africa}, on the other hand, has a high spatial resolution.
However, it is a rolling-horizon optimal dispatch model, and does not optimise capacity expansion.

We conclude that there is currently no large-scale high resolution model of the African energy system addressing the capacity expansion problem.

\begin{figure}[h!]
\centerline{\includegraphics[trim={0cm 0.2cm 0cm 3cm},clip,width = 0.65\textwidth]{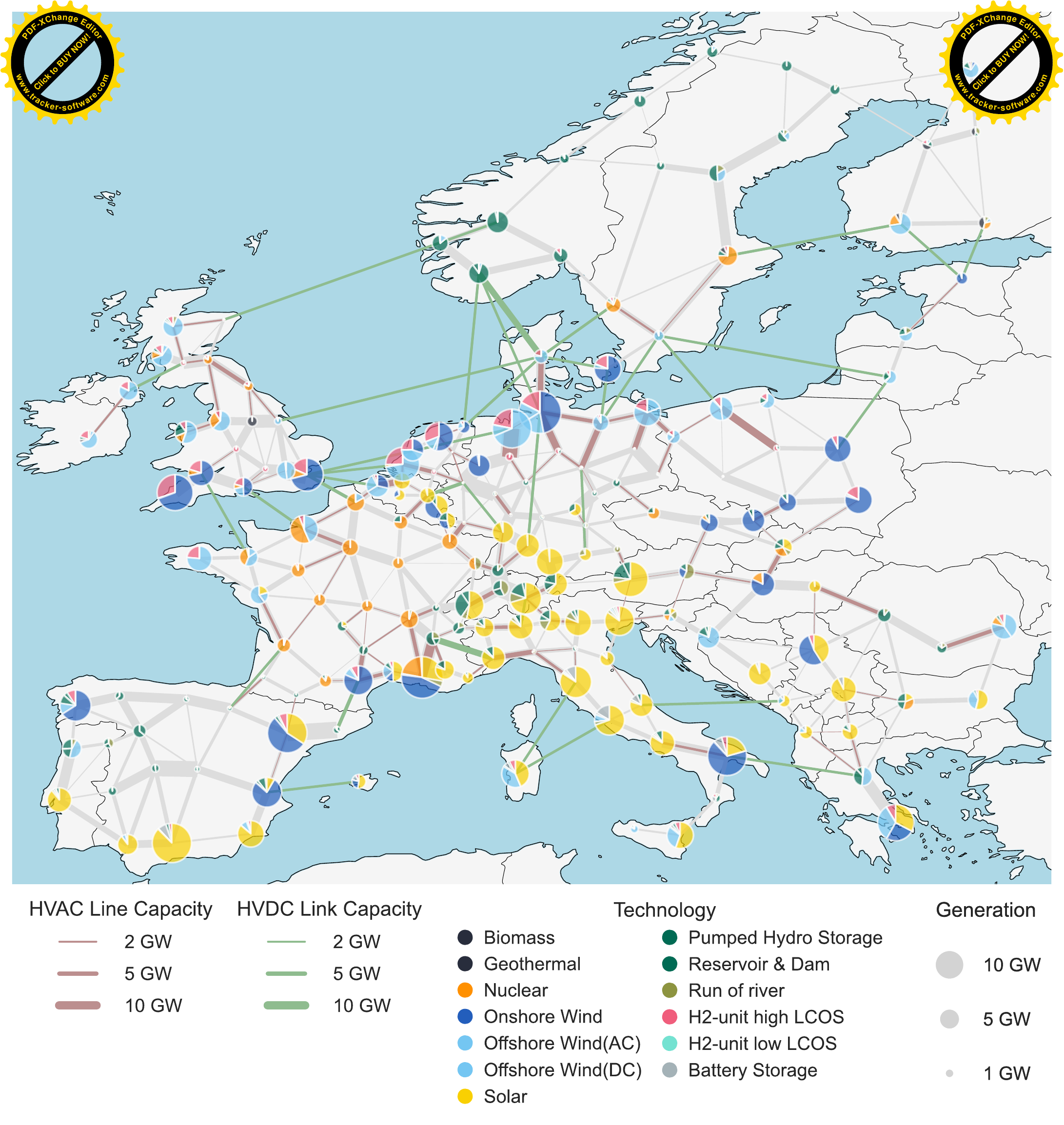}}
\caption{A network simulation from \emph{PyPSA-Eur} that shows the planned achievements in the future African model. Map showing optimal generation, storage and network expansion in Europe under a \SI{100}{\percent} emission reduction scenario and technology data for 2030. Light grey lines showing the existing installed network capacity~\cite{parzen-neumann-ea-2021}.}
\label{EU-Map}
\end{figure}

\section{Roadmap to an Open Energy System Model for Africa}

The \emph{PyPSA meets Africa} project aims to construct the first high resolution open model of the African energy system capable of optimal capacity expansion. We are building an automated, modular and reproducible workflow based on \emph{PyPSA} (Python for Power System Analysis).

\emph{PyPSA}~\cite{brown-horsch-ea-2018} is an open modelling framework with a significant number of active users.
It is used for capacity expansion and dispatch optimisation of energy systems.
A principal feature of \emph{PyPSA} is the option for high-resolution modelling of optimal power flow in transmission networks.
This makes \emph{PyPSA} well suited to large-scale models including conventional and renewable energy sources, storage technologies as well as energy carriers including natural gas and hydrogen.

A prominent example for a \emph{PyPSA}-based model is \emph{PyPSA-Eur}~\cite{PyPSAEur}, a high resolution model of the European electricity
system on a transmission grid level.
\autoref{EU-Map} shows the \emph{PyPSA-Eur} model applied in a capacity expansion scenario for an European electricity
system with \SI{100}{\percent} emission reduction in 2030.
It provides a vision of the possible results in the African model based on \emph{PyPSA}.

The model under development by \emph{PyPSA meets Africa} builds on the capability of \emph{PyPSA-Eur}.
It will be the first model of the African transmission grid at a high, adjustable spatial resolution, capable of optimal capacity expansion.
Moreover, this is done with a modular workflow based on open data meeting the FAIR principles~\cite{wilkinson-dumontier-ea-2016}.
As a result, the model can easily be updated with improved data as it becomes available.

The project aims to work as much as possible with existing open source libraries and working code bases; see also the later subsections.
This will allow us to couple the African model with \emph{PyPSA-Eur}, enabling cross-beneficial source code development, intercontinental studies, and a larger user base.
Additionally, we foresee to add sector-coupling to the African model in a similar way as to \emph{PyPSA-Eur-Sec}.
The ultimate goal is to create a long-term maintained and supported open model that is useful for industry, research and policy purposes.

The project is split into multiple work packages which are layed out in the next subsections including an outlook on the methodology planned for each work package:

\begin{enumerate}
    \item Demand modelling
    \item Conventional generator modelling
    \item RES modelling
    \item Land coverage constraint modelling
    \item Network and substation modelling
    \item Data creation and validation 
\end{enumerate}

\subsection{Demand modelling} 
In order to model the electricity consumption, GlobalEnergyGIS (GEGIS) will be employed to obtain a hourly time-series of demand.
GEGIS applies a machine learning (ML) approach based on existing electricity demand time-series data, population densities and spatially resolved income data.
This data is complemented with scenario information from the Shared Socioeconomic Pathways (SSP) and weather data from ERA5\footnote{\url{{https://www.ecmwf.int/en/forecasts/dataset/ecmwf-reanalysis-v5}}} (i.e. hourly estimates of atmospheric and oceanic climate variables) to generate hourly outputs for arbitrary regions until 2050.

\subsection{Existing generator modelling} 
Regarding the modelling of existing generators, data is made available through the Global Power Plant Database (GPPD)~\cite{globalenergyobservatory-google-ea-2018}, which is one of the principal open global power plant databases.
Commercial power plant databases with restricted access historically were dominant in policy-making and research, but in recent years open databases are closing the gap.
For conventional generation (i.e. coal, natural gas, oil, hydro), the GPPD has a coverage of 80--100\% (depending on the category) in terms of installed capacity, when compared to the state-of-the-art commercial sources~\cite{byers-friedrich-ea-2019}.

The database can be integrated with \emph{PyPSA} using the \emph{powerplantmatching} tool~\cite{gotzens-heinrichs-ea-2019}.
The latter also enables integration with other databases.
As such, our model can incorporate new and improved datasets as they become available.

Hydropower plays an important role in Africa, and is expected to continue doing so.
While the GPPD includes data on existing hydro plants, we can also use existing geospatial analyses~\cite{korkovelos-mentis-ea-2018} of the potential for new hydro plants. Additionally, it is also necessary to extract the time-series of hydrological inputs. These define the operation of hydropower plants and can be derived using the LISFLOOD model \cite{pavicevic-defelice-ea-2021}.

\subsection{RES generator modelling} 
In terms of RES modelling, the project mainly focuses on new installation potential. As widely known, weather data such as wind speed/direction and solar irradiation play an important role in RES modelling. Using \emph{atlite}~\cite{atlite} package, spatially resolved potentials and hourly time-series of wind and solar data are modelled into generation output from different plants. 

Currently, \emph{atlite} supports wind turbines and solar PV as generators and uses ERA5 complemented with irradiation data from SARAH2\footnote{\url{{https://wui.cmsaf.eu/safira/action/viewDoiDetails?acronym=SARAH_V002}}}.

The project aims to increase the variety of RES models to include concentrated solar power and marine energy like wave or tidal generators.

\subsection{Land coverage constraint modelling} 
Different types of land use, extreme terrain conditions and protected areas constrain the eligible areas for RES development, placement of transmission lines, etc.
Both land and sea constraints can be integrated in \emph{atlite}.
In addition to the oceanic climate factors, water depth is a constraint for offshore and marine energy turbines where bathymetric dataset will be obtained from General Bathymetric Chart of the Oceans (GEBCO)\footnote{\url{{https://www.gebco.net/data_and_products/gridded_bathymetry_data/}}}. Protected areas such as national parks will be researched and integrated into the model which provides an aspect of novelty. Population distribution and land cover surveys such as Corine Land Cover can be used to indicate exclusion regions like regions with high population densities, farmland or extreme terrain.

\subsection{Network and substation modelling}
The network topology for Africa will mainly be extracted from a dataset compiled by the World Bank Group, which is based on a number of different sources ~\cite{arderne-2017}. The dataset includes 1818 inputs for transmission lines (above 110 kV) with a total length of 4204 km. OpenStreetMap can be alternatively used, however, might require more data cleaning to be reliable. Both datasets provide voltage level and number of lines as physical information. To allow power flow studies the physical information needs to be included. Therefore, similar to \emph{PyPSA-Eur}~\cite{PyPSAEur}, typical physical parameters such as length dependent resistance and reactance, current thermal level and apparent power limit will be added.

\begin{figure}[h]
\centerline{\includegraphics[trim={4cm 10cm 4cm 9cm},clip,width = 0.7\textwidth]{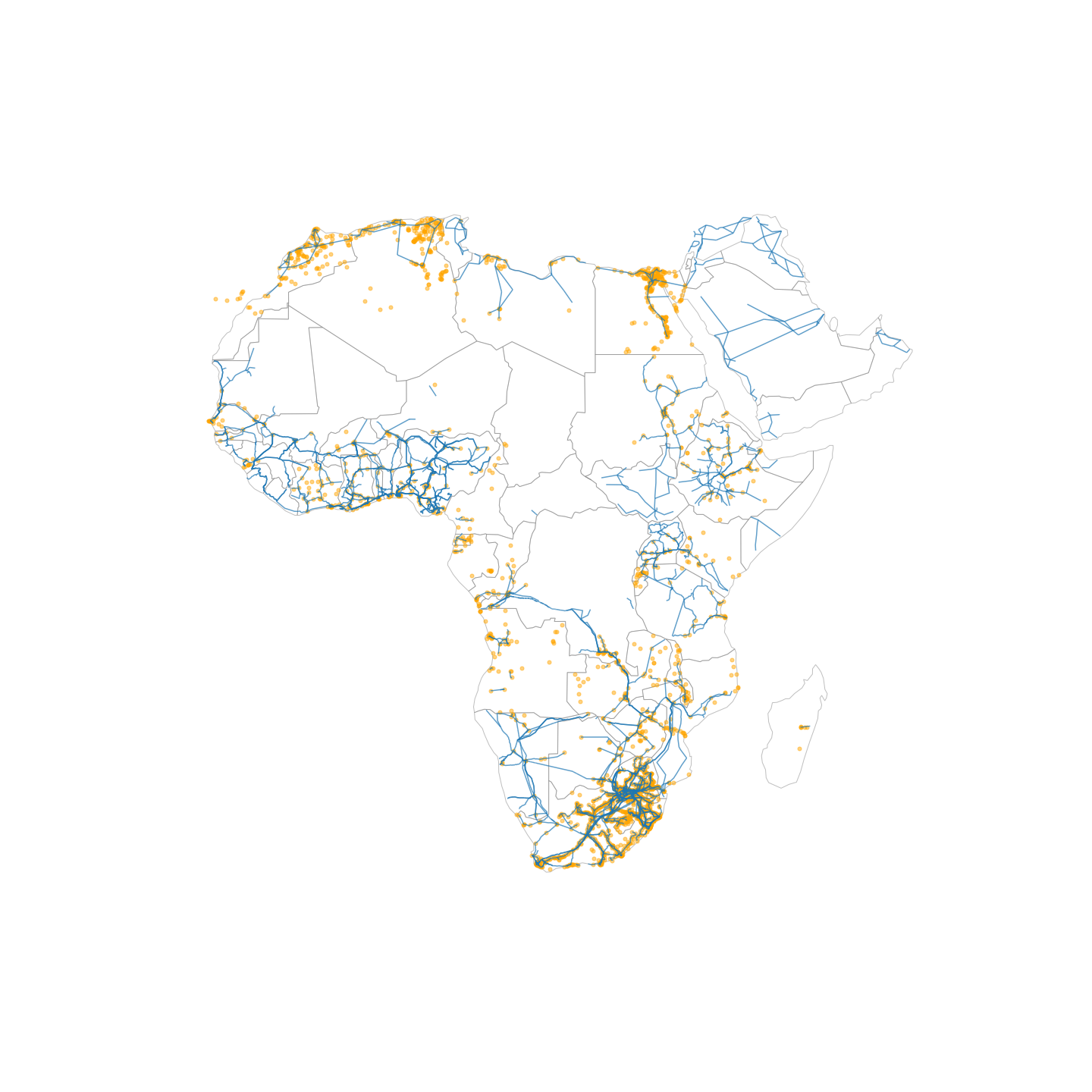}}
\caption{The African map shows numerous transmission lines and substations. Blue lines show both the planned and existing transmission lines of over 110 kV rating in Africa and the MENA region. The points in orange denote the existing substations. Data are available at \url{https://energydata.info/dataset} and OpenStreetMap}.
\label{AfricaMApTransmission}
\end{figure}

The substation geolocation with voltage level will be extracted from \emph{Open Street Maps} (OSM). To be achieve computation on typical 16GB RAM computers, a relatively  new efficient OSM extraction package \emph{esy-osmfilter} will be applied that is four times faster than the common alternative, \emph{OSMOSIS} ~\cite{pluta-lunsdorf-2020}. Initial results of the OSM substation extraction process (see \cref{AfricaMApTransmission}) show 2721 transmission substation data points for Africa.

After collecting the substation and line data, it is essential to build a consistent network format that can be used by existing \emph{PyPSA-Eur} scripts. This requires approaches of open software packages such as \emph{GridKit} \footnote{\url{https://github.com/PyPSA/GridKit}} and \emph{OpenGridMap}\footnote{\url{https://github.com/OpenGridMap/transnet}}, both building automatic inferences from mapped grid data.

As for \emph{PyPSA-Eur}, the country shapes and bus location will be used to split areas into Voronoi cells which are assumed to be connected to low voltage networks. They define the spatial resolution of the simulation. Each of the Voronoi cells contains information on existing power plant capacities, renewable resource potential timeseries, as well as demand timeseries at the substation location. The spatial model resolution can be changed by aggregating the cells by k-means clustering. Altering the spatial resolution is beneficial for reducing the computational expense for simulations that do not require the same level of detail. For instance, while a high resolution optimisation for Africa with up to 1000 nodes/regions might be required for a detailed continental study, a one-node per country resolution can be sufficient for simplified tests.

\subsection{Data creation and validation} 

The amount of data on energy assets might vary per country (see \cref{AfricaMApTransmission}). The missing data can be either added by local energy authorities or using asset recognition techniques on satellite imagery. It is known that data from centralised organisations such as the European Network of Transmission System Operator (ENTSO-E) may be out-of-date or unreliable \cite{gotzens-heinrichs-ea-2019,PyPSAEur}. Therefore, ML techniques can be applied to both validate existing data and create new data from satellite images.

In the context of the electricity grid applications, ML can be used to create missing data and increase manual validation speed by at least 33-fold compared to manual mapping strategies~\cite{developmentseed-2018}. 

Using ML techniques, high voltage lines, substations~\cite{developmentseed-2018}, solar PV \cite{dehoog-maetschke-ea-2020} and wind turbines~\cite{zhou-irvin-ea-2019} were identified from satellite imagery. This involves creating a training data set which consists of marking energy assets on the images. Once the algorithm is trained, it is able to identify the chosen energy asset in several other images. The information gained from ML can be, for instance, the area and inclination of PV arrays that may be used to estimate their annual energy output. Other physical parameters such as voltage level of substations and age of assets can be also extracted by training ML on specific substation data and analysing satellite images over multiple years.

The \emph{PyPSA meets Africa} project prioritises the creation of validated substation and high voltage line datasets which will be made publicly available in line with the open approach. There are plans to use ML for the identification of other energy assets in the future.

\section{Conclusions}
In this paper, we discussed the importance of open energy systems modelling in the African continent. The existing work in terms of both academic literature, technical reports and initiatives were reviewed. Due to the increasing number of projects that promote interconnection and RES integration, we identified the gap in the literature for a centralised continent-level model for Africa. 
To address this point, we presented our proposed roadmap to \emph{PyPSA meets Africa} which is an open energy modelling and data sharing initiative for Africa open to a wider audience. The developments of the new open python-based investment and dispatch optimisation model for the African energy system will have a high spatial and temporal resolution and detailed modelling of the existing and planned transmission grid.
The aim of the initiative is to create an open energy system model for Africa and provide long-term support, maintenance and development. The model allows exploratory planning studies such the integration of renewable energy sources and other technologies such as grid-scale batteries that can be explored for a more resilient and sustainable African electricity grid~\cite{battery} via ancillary services. This open modelling initiative is expected to accelerate the model developments, empowers research within  Africa and hence, the integration of renewable energy technologies. 

The proposed innovation could potentially contribute to the industrialisation of Africa in accordance with the UN's SDGs, the Paris Agreement and Africa´s Agenda 2063.
The plans for the future work include use of machine learning techniques in order to form new energy datasets and validate the existing ones~\cite{developmentseed-2018}, add sector-coupling capabilities to the African model (as already done in PyPSA-Eur-Sec \cite{PyPSAEurSec}) and finally link the African model to the European. As result, model improvements in Africa can benefit Europe and the vice versa.

\section*{Acknowledgments}
This research is supported by the Engineering and Physical Sciences Research Centre (EPSRC) grants: EP/R513209/1 for Doctoral Training Partnership, EP/P007805/1 for the Centre for Advanced Materials for Renewable Energy Generation (CAMREG) and EP/P001173/1 for Centre of Energy System Integration (CESI).

\bibliographystyle{IEEEtran}

\bibliography{bibliography}

\end{document}